\documentclass[%
reprint,
superscriptaddress,
showpacs,
amsmath,amssymb,
aps,
prl,
longbibliography,
]{revtex4-2}

\usepackage{psfrag,graphicx,epsfig,color}
\usepackage{dcolumn}
\usepackage{bm}
\usepackage{natbib}
\usepackage{float}
\usepackage[usenames,dvipsnames,svgnames,table]{xcolor}
\usepackage{subfigure}
\usepackage{nicefrac}

\def\re    {R_\lambda}

\definecolor{mygreen}{rgb}{0,0.7,0.}

\begin{document}

\title{
Saturation and multifractality of Lagrangian and Eulerian \\ scaling exponents 
in 3D isotropic turbulence  
}

\author{Dhawal Buaria }
\email[]{dhawal.buaria@nyu.edu}
\affiliation{Tandon School of Engineering, New York University, New York, NY 11201, USA}
\affiliation{Max Planck Institute for Dynamics and Self-Organization, 37077 G\"ottingen, Germany}
\author{Katepalli R. Sreenivasan}
\affiliation{Tandon School of Engineering, New York University, New York, NY 11201, USA}
\affiliation{Department of Physics and the Courant Institute of Mathematical Sciences, New York University,
New York, NY 10012, USA}

\date{\today}

\begin{abstract}

Inertial range scaling exponents for both  
Lagrangian and Eulerian structure functions are obtained from
direct numerical simulations of isotropic
turbulence in triply periodic domains at Taylor-scale Reynolds number
up to 1300. We reaffirm that transverse Eulerian scaling exponents
saturate at $\approx 2.1$ for moment orders $p \ge 10$,
significantly differing from the longitudinal exponents
(which are predicted to saturate at $\approx 7.3$ for $p \ge 30$ from a 
recent theory).
The Lagrangian scaling exponents likewise saturate at $\approx 2$ for $p \ge 8$.
The saturation of Lagrangian exponents and transverse Eulerian  exponents
is related by the same multifractal spectrum 
by utilizing the well known frozen hypothesis
to relate spatial and temporal scales.
Furthermore, this spectrum is different
from the known spectra for Eulerian longitudinal exponents,
suggesting that that Lagrangian intermittency is characterized 
solely by transverse Eulerian intermittency.
We discuss possible implication of this outlook when extending
multifractal predictions to the dissipation range, 
especially for Lagrangian acceleration.

\end{abstract}

\maketitle


Turbulent flows in nature and engineering comprise
a hierarchy of eddies, with smaller eddies
coexisting within larger ones and extracting energy from them.
To understand the deformation and rotation of smaller eddies, the
key mechanisms driving energy transfers, it is essential
to examine the velocity increments across a smaller eddy of
size $r \ll L$ (say), where $L$ is the large-eddy size
\cite{K41, MY.II, Frisch95}. 
The longitudinal velocity increment $\delta u_r = u(x+r) - u(x)$ 
corresponds to the case when the velocity component $u(x)$ 
is in the direction of separation $r$. 
For velocity $v(x)$ taken orthogonal to $r$, 
transverse velocity increment $\delta v_r = v(x+r) - v(x)$
is obtained.

The motivation to study the small eddies (and hence 
velocity increments) 
stems from their purported universality, postulated
by Kolmogorov (1941) \cite{K41}---K41 henceforth---which
has since become the backbone of turbulence
theory and modeling \cite{Frisch95, popebook}. 
Building upon K41, one surmises that 
moments of increments
$\langle (\delta u_r)^p \rangle $, called 
structure functions, follow a universal power-law scaling
in the so-called inertial-range:
\begin{align}
    S_p(r) \equiv \langle (\delta u_r)^p \rangle \sim  r^{\zeta_p} \ , 
    \ \ \ \ \eta \ll r \ll L \ ,
\end{align}
where $\eta$ is the viscous cutoff scale. 
Establishing such a simple scaling
enables dramatic simplification in studying
a wide range of turbulent flows, and thus, 
structure functions have been of persistent interest
and a cornerstone of turbulence theory
\cite{K62, MY.II, Frisch95, Sreeni97}. 
K41 originally postulated  $\zeta_p = p/3$;
this result is known to be exact
for $p=3$, i.e., $\zeta_3=1$, but extensive studies
from \cite{vanatta} to \cite{iyer20} (and others in between)
have clearly established nonlinear deviations
of $\zeta_p$ from $p/3$ for $p\ne3$.
This so-called anomalous scaling is attributed to the intermittency 
of interscale energy transfer processes 
(see, e.g., \cite{K62, MY.II, Frisch95, Sreeni97}).

Since turbulence can also be fundamentally explored
from a Lagrangian viewpoint
\cite{MY.II, wyngaard, sawford.2001, falkovich01, toschi:2009},
forceful arguments can be similarly made for
Lagrangian velocity increments $\delta u_\tau = u(t + \tau) - u(t)$ 
over time lag $\tau$, measured along fluid-particle trajectories, and
Lagrangian structure functions $\langle |\delta u_\tau |^p \rangle $
defined therefrom 
\footnote{Absolute value is taken 
for Lagrangian increments since their odd moments
are zero}. 
Extension of K41 phenomenology to
Lagrangian increments gives:
\begin{align}
    S^L_p (\tau) \equiv \langle |\delta u_\tau|^p \rangle \sim  \tau^{\zeta_p^L} \ , 
    \ \ \ \ \tau_\eta \ll \tau \ll T_L
\end{align}
where the temporal inertial-range is defined using $T_L$, the 
Lagrangian integral time
and $\tau_\eta$, the time-scale of viscous dissipation \cite{MY.II}. 
Since Lagrangian trajectories trace the underlying
Eulerian field, it is natural to expect that a relation between Lagrangian 
and Eulerian exponents can be obtained.

Using K41, one obtains $\zeta_p^L=p/2$ \cite{MY.II};
but, experimental and numerical studies again show 
nonlinear deviations from this prediction
\cite{Sawford03, mordant2004, Xu06, sawford15}.
Several attempts have been made 
\cite{borgas93, biferale2004, arneodo} to 
quantify these deviations in terms of Eulerian intermittency,
but they remain deficient
for at least two reasons. First, the temporal
scaling range in turbulence is substantially more restrictive
than spatial scaling range \cite{MY.II, Frisch95}, making
it difficult to robustly extract the Lagrangian scaling exponents.
Second, past attempts have overwhelmingly focused on
characterizing Lagrangian
intermittency from longitudinal Eulerian intermittency,
assuming that longitudinal and transverse exponents are identical, 
despite counter-evidence
\cite{dhruva97, chen97, gl97, shen02, gotoh02, grauer2012}. 

In this Letter, presenting new data from direct numerical simulations (DNS)
of isotropic turbulence at higher Reynolds numbers, we address both these
challenges. 
We extract both Lagrangian and Eulerian scaling exponents.
Our Eulerian results reaffirm recent results \cite{iyer20}. 
We then demonstrate an excellent correspondence 
between Lagrangian exponents and transverse Eulerian exponents, 
using as basis the same multifractal spectrum;
this is different from the multifractal spectrum
for longitudinal exponents, whose use in the past has failed to explain 
Lagrangian intermittency \cite{BS_PRL_2022, Sawford03, mordant2004, Xu06, sawford15}).


\paragraph*{Direct Numerical Simulations:}
The description of DNS is necessarily brief here because they
have been described 
in many recent works
\cite{BS2020, BBP2020, BP2021, BPB2022, BS2022, bs_pnas_23}.
The simulations correspond to the canonical setup of 
forced stationary isotropic turbulence in a triply periodic domain 
and are carried out
using the highly accurate Fourier pseudo-spectral methods in space
and second-order Runge-Kutta integration in time;
the large scales are numerically forced
to achieve statistical stationarity \cite{Ishihara09, Rogallo}.
A key feature of the present data is that we have achieved
a wide range of Taylor-scale Reynolds number $\re$,
going from $140-1300$ (on grids of up to $12288^3$ points)
while maintaining excellent small-scale
resolution \cite{BPBY2019, BBP2020}. 
For Lagrangian statistics, a large population of fluid 
particles is tracked together with the Eulerian field.
For $\re \le 650$, up to 64M particles are tracked
for each case, whereas for $\re = 1300$, 256M particles are tracked
(with M$=1024^2$) \cite{BSY.2015, BYS.2016, buaria.cpc}, providing
ample statistics for convergence. 

\begin{figure}
\centering
\includegraphics[width=0.48\textwidth]{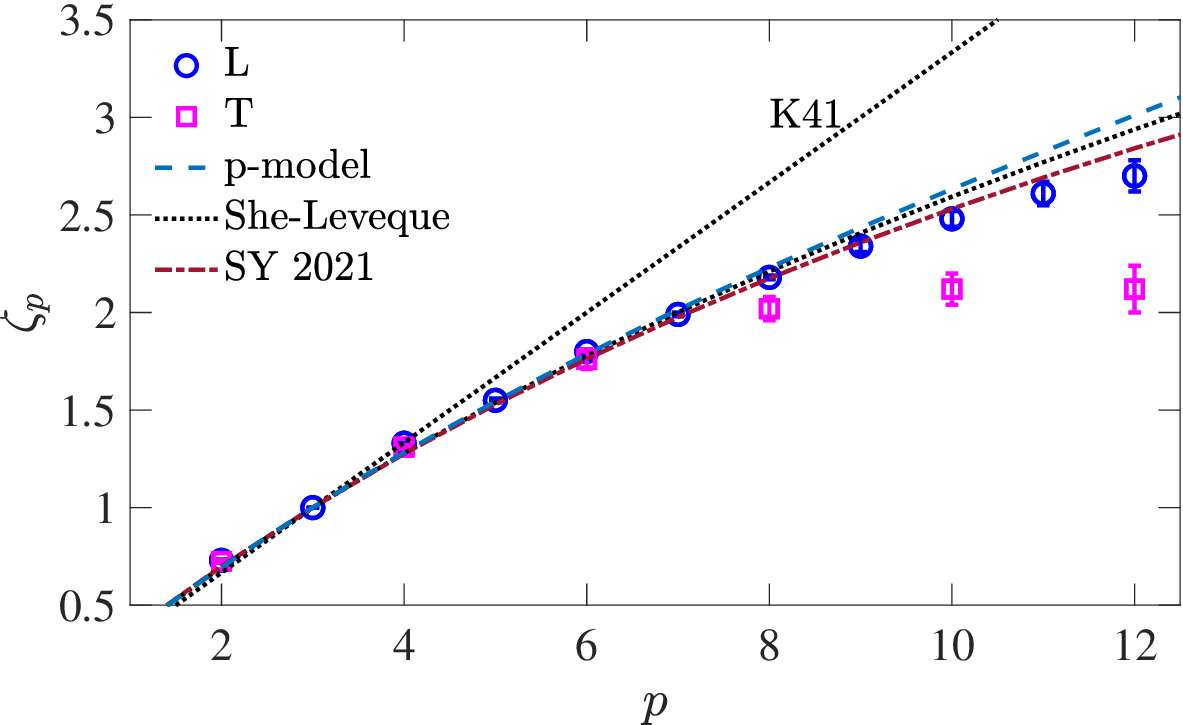} 
\caption{
Inertial-range scaling exponents for longitudinal 
and transverse Eulerian structure functions,
the former from \cite{iyer20, SY2021} and the latter 
from the present data (consistent with \cite{iyer20}). 
Various theoretical predictions
\cite{K41, MS87, SL94, SY2021} are also shown.
The transverse exponents depart from all predictions and saturate.
}
\label{fig:eul}
\end{figure}

\paragraph*{Saturation of transverse exponents:} 
Anomalous scaling confers upon each moment order 
a separate and independent significance, instead of a mutual
dependence (such as $\zeta_p = p/3$ based on K41). 
Multifractals have enjoyed considerable success in describing this behavior
\cite{Frisch95, Sreeni97},
but lack 
any direct connection to Navier-Stokes equations. 
Further, recent DNS at high $\re$ 
have shown noticeable departures
of $\zeta_p$ from multifractal predictions for high orders \cite{iyer20}. 
Instead, starting from Navier-Stokes equations, 
a recent theory \cite{SY2021} was able to 
provide an improved prediction for $\zeta_p$.
Additionally, this theory also predicts that  
longitudinal exponents saturate with the moment-order, 
i.e., $\lim\limits_{p\to\infty }  \zeta_p \to \text{constant}$.  

Recall that the
transverse exponents are defined by the relation 
$S_p^{tr} \sim r^{\zeta_p^{tr}}$, where
$S_p^{tr} (r) \equiv \langle |\delta v_r|^p \rangle $. 
(Absolute values are taken as the odd-moments
are zero from symmetry.) 
Multifractal models based on phenomenological considerations
do not differentiate 
between longitudinal and transverse exponents, i.e. $\zeta_{2p}^{tr} = \zeta_{2p}$, 
and general arguments have also been advanced to the same end 
\cite{lvov97, Nelkin90}.
However, several studies 
have persistently pointed out that the two sets of exponents are different 
\cite{dhruva97, chen97, gl97, shen02, gotoh02, grauer2012}; recent work
at high $\re$ \cite{iyer20} has confirmed the differences, also showing that
transverse exponents saturate: $\zeta_\infty^{tr} \approx 2.1 $ for $p\ge10$. 
Incidentally, this saturation is very different from 
$\zeta_\infty \approx 7.3$ (for $p \ge 30$) 
predicted for longitudinal exponents in \cite{SY2021}.

\begin{figure}
\centering
\includegraphics[width=0.48\textwidth]{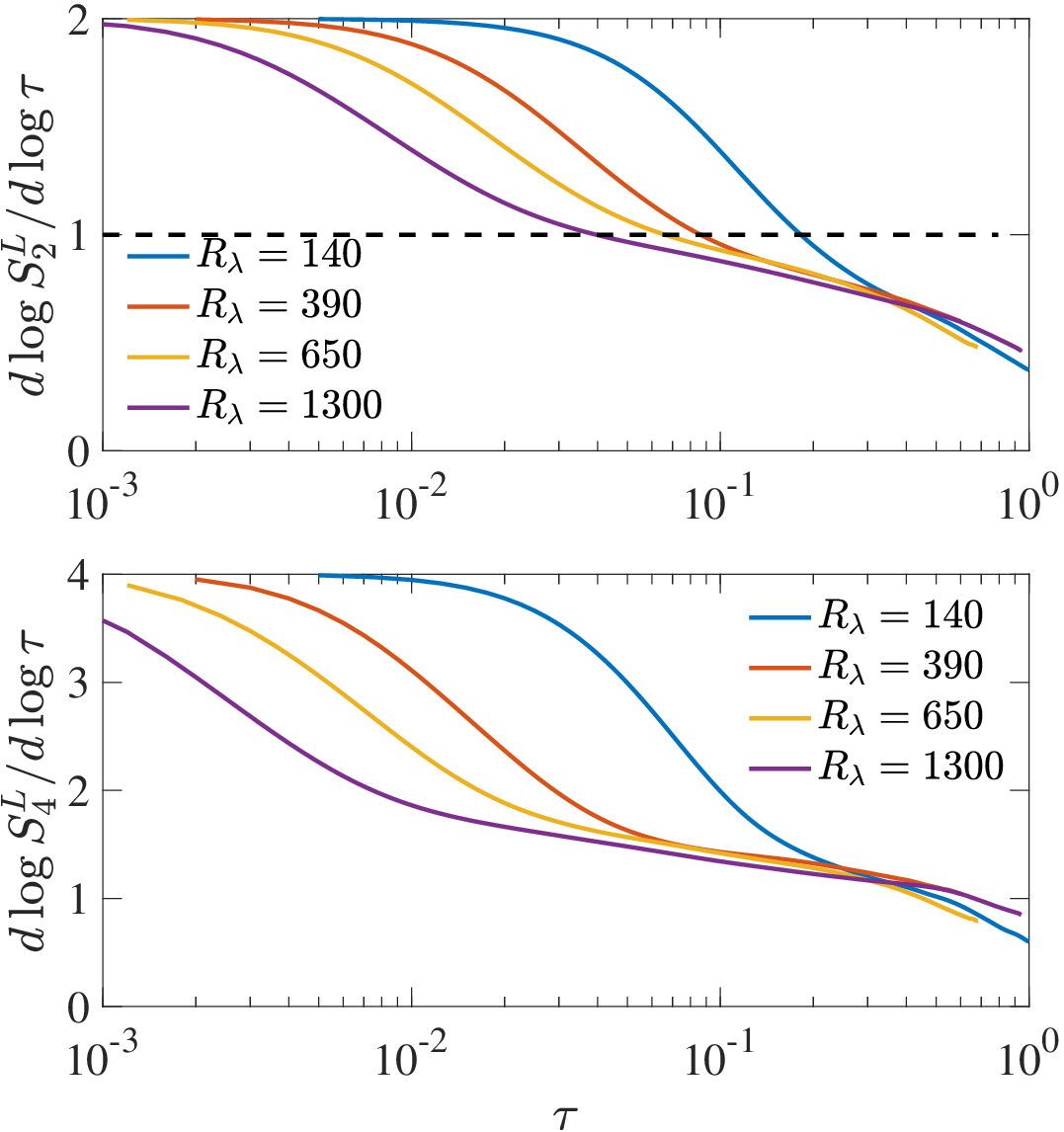} 
\caption{
Local slopes for (a) second and (b) fourth-order Lagrangian
structure functions at various $\re$.
}
\label{fig:ls}
\end{figure}

These findings are summarized in Fig.~\ref{fig:eul}, 
showing the longitudinal and transverse exponents.
Also included are K41 prediction, multifractal results \cite{MS87, SL94} 
and the result from \cite{SY2021}. Important considerations go into establishing 
the reliability of high-order exponents with respect to 
statistical convergence, 
adequacy of grid resolution, and $\re$-dependence.
This discussion can be found in
\cite{iyer20} and will not be repeated here. 
Instead, we focus on $\zeta_p^{tr}$, 
which clearly depart 
from $\zeta_p$ and saturate for $p\ge 10$. 
The implication of 
different longitudinal and transverse exponents 
for small-scale universality  
is discussed later; we first 
demonstrate how $\zeta_p^{tr}$ is directly related to the Lagrangian
exponents.

\begin{figure}
\centering
\includegraphics[width=0.48\textwidth]{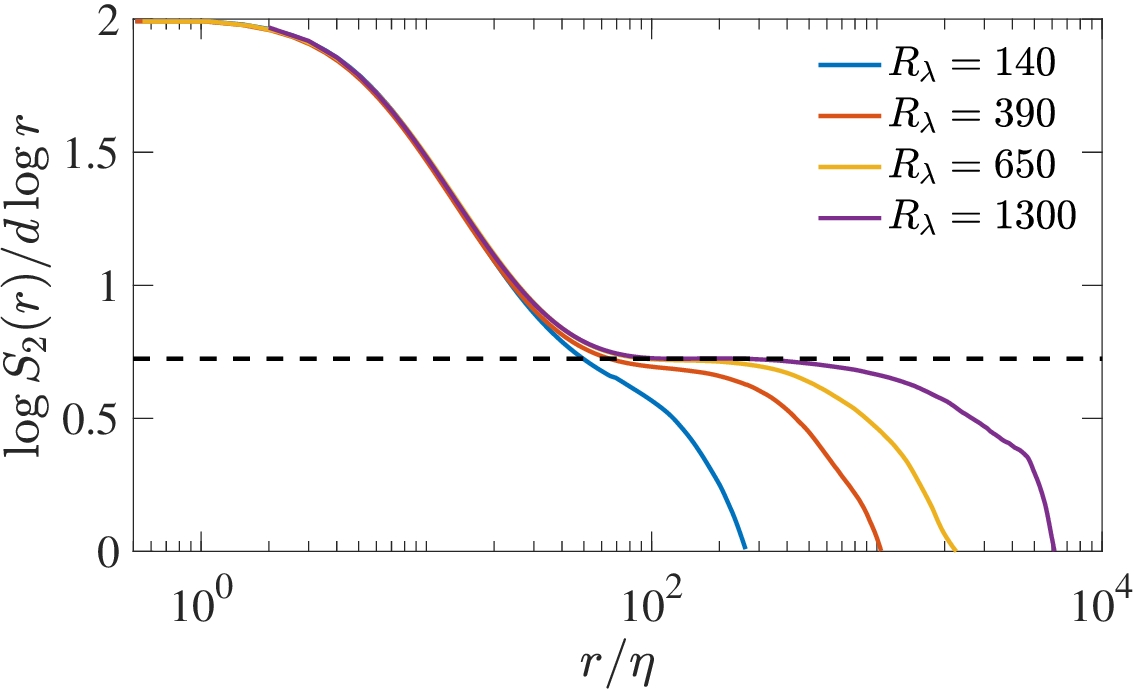} 
\caption{
Local slopes for the Eulerian second-order structure functions
at different $\re$. In contrast to 
Lagrangian data in Fig.~\ref{fig:ls}, a clear inertial-range emerges 
with Reynolds number.
}
\label{fig:eul_ls}
\end{figure}

\paragraph*{Lagrangian exponents from DNS:}
Robust extraction of scaling exponents requires sufficient
scale separation to allow a proper inertial-range to 
exist.  
The Eulerian spatial scale separation 
for the highest $\re=1300$ is $L/\eta \approx 2500$ \cite{iyer20},
while the temporal range
is $T_L/\tau_K \approx 105$ \cite{buaria.thesis}, 
thus making it inherently difficult to obtain a proper
Lagrangian inertial-range \cite{sawford2011, buaria.com}. 
This difficulty is highlighted
in Fig.~\ref{fig:ls}, which shows
the log local slope of $S_p^L(\tau)$
at various $\re$, 
for $p=2$ and $4$ in panels (a) and (b), respectively; 
although there is a suggestion of a plateau for the fourth-order, 
the local slopes of the curves
are still changing with $\re$. This is in contrast to the corresponding 
Eulerian result for $p=2$, 
shown in Fig.~\ref{fig:eul_ls}, 
where a clear inertial-range emerges with $\re$.

\begin{figure}
\centering
\includegraphics[width=0.48\textwidth]{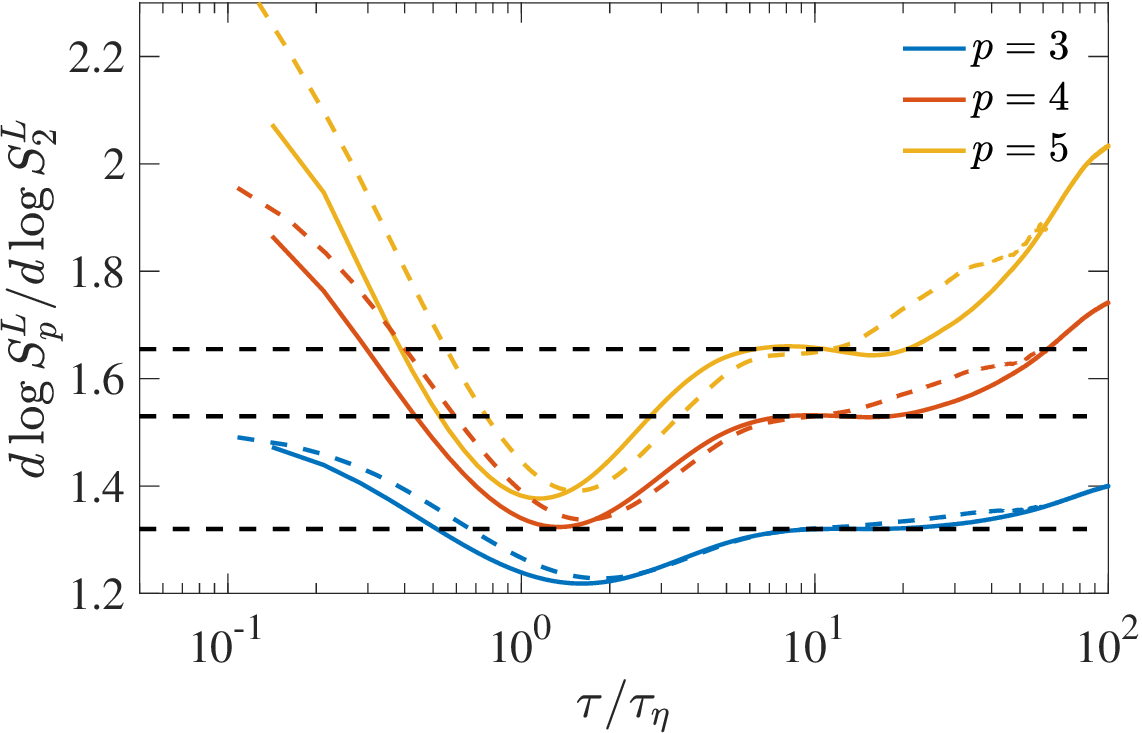} 
\caption{
Ratio of local slope for 
$p$-th order Lagrangian
structure function to second-order, 
for $p=3-5$, 
at $\re=1300$ (solid lines) 
and $\re=650$ (dashed lines). 
}
\label{fig:ls2}
\end{figure}

Because of this difficulty, Lagrangian exponents cannot
be directly extracted even at the highest $\re$ available.
However, by using extended self-similarity \cite{benzi93}, 
we can obtain the exponents  
with respect to the second-order 
\cite{sawford15}. 
Fig.~\ref{fig:ls2}  shows
the ratio of local slope
of $S^L_p(\tau)$ to that of $S^L_2(\tau)$. Evidently,
a conspicuous plateau emerges for different orders 
in the same scaling range, seemingly independent of $\re$.
Thus, we can extract the ratios $\zeta^L_p/\zeta^L_2$,
which also was the practice in earlier works 
\cite{mordant2004, Xu06, sawford15}.
The justification for using
$\zeta_2^L$ as the reference 
comes from the expectation  
$S_2^L \sim \langle \epsilon \rangle \tau$ \cite{MY.II};
since the mean dissipation appears linearly, 
the result $\zeta_2^L=1$ is free of intermittency
(akin to $\zeta_3=1$ for Eulerian 
exponents \cite{K41b}). 

\begin{figure}
\centering
\includegraphics[width=0.48\textwidth]{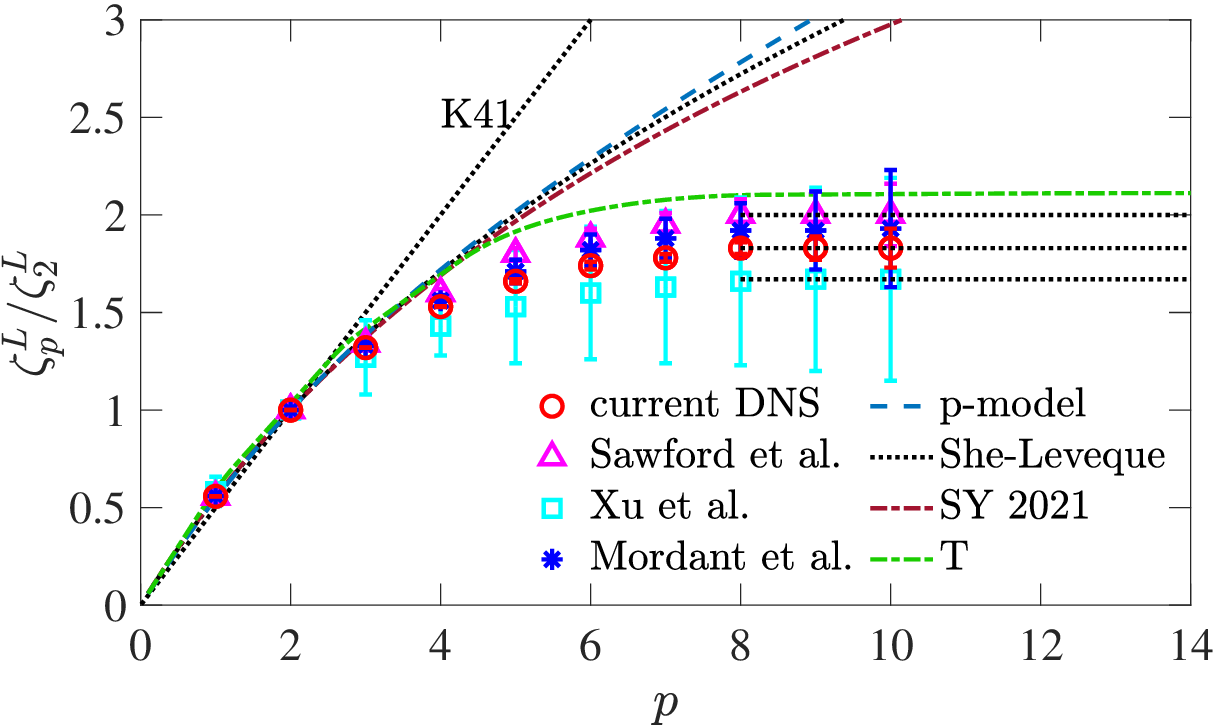} 
\caption{
Lagrangian scaling exponents and comparison with 
prior results and various 
multifractal models. The prediction from the transverse exponents is
shown by the green curve that saturates for large $p$.
}
\label{fig:lag}
\end{figure}

Extending the procedure 
in Fig.~\ref{fig:ls2}, 
the ratios 
$\zeta^L_{p}/\zeta^L_2$ are extracted
for upto $p=10$ and shown in Fig.~\ref{fig:lag}. 
We also include earlier results from both experiments and DNS 
\cite{arneodo, mordant2004, Xu06, sawford15}, obtained
at comparatively lower $\re$.
Overall, the current results at higher $\re$
are in excellent agreement with prior results
(which had larger error bars). 
A remarkable result, endemic to all
cases, is that 
the Lagrangian exponents saturate 
for $p \gtrsim 8$, similar to
the transverse Eulerian exponents
in Fig.~\ref{fig:eul}. 
The data in Fig.~\ref{fig:lag} are also 
compared with various
predictions, which we discuss next.

\paragraph*{The multifractal framework:}
Evidently, the data in Fig.~\ref{fig:lag} strongly deviate from K41.
Following \cite{Frisch95,biferale2004}, we will consider the well 
known multifractal model
for relating Eulerian and Lagrangian exponents. 
The key concept in multifractals 
is that the (Eulerian) velocity increment $\delta u_r$
over a scale $r$ is H\"older continuous, i.e., 
$\delta u_r \sim r^h$, where $h$ is the local
H\"older exponent with the multifracal
spectrum $D(h)$ \cite{benzi1984, Frisch95}. From this local scaling,
Eulerian structure functions
are readily derived by integrating over all 
possible $h$, as
$\langle (\delta u_r)^p \rangle \sim \int_h r^{ph + 3 - D(h)} dh $. 
Using steepest-descent argument
for $r \ll L$ gives
\begin{align}
\zeta_p = \inf_h \left[ ph + 3 - D(h) \right ] \ .
\label{eq:zeul}    
\end{align}

The Lagrangian extension of multifractals
relies on the phenomenological assumption that 
spatial and temporal separations are interchangeable: 
$r \sim \tau \delta u_r$, 
akin to frozen flow hypothesis,
with $\delta u_r \sim \delta u_\tau$ 
\cite{biferale2004}.
This stipulation gives $\delta u_\tau \sim \tau^{h/(1-h)}$, 
resulting in
the Lagrangian exponents 
\begin{align}
\zeta_p^L = \inf_h \left[ \frac{ph + 3 - D(h)}{1-h} \right ] \ .  
\label{eq:zlag}   
\end{align}
Thus, Lagrangian exponents can be directly
predicted using the Eulerian multifractal spectrum $D(h)$.
Since past works have predominantly focused on 
Eulerian longitudinal exponents, with the implicit assumption
that transverse exponents are same, the 
$D(h)$ of the longitudinal exponents 
has been used to infer Lagrangian exponents.
However, such
predictions do not work as we see next.

\begin{figure}[h]
\centering
\includegraphics[width=0.48\textwidth]{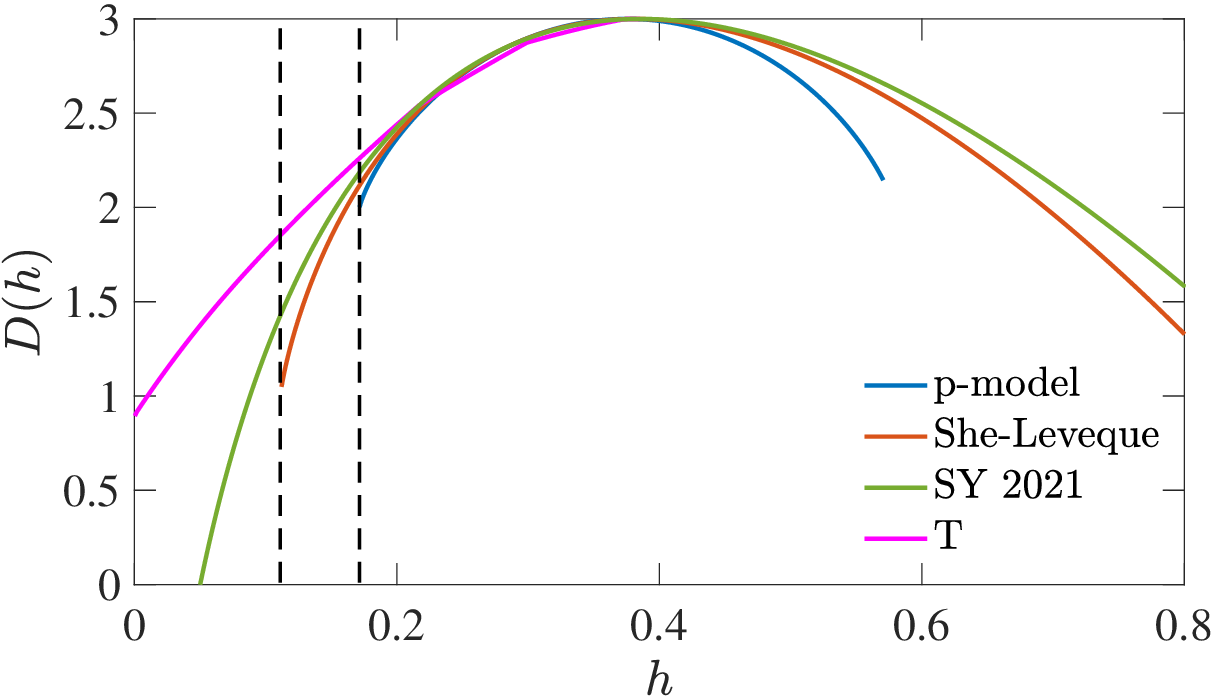} 
\caption{
The multifractal spectra for various models.
The vertical dashed lines at $\frac{1}{3}\log_2 (0.7) (\approx 0.17)$ 
and $\frac{1}{9}$ mark the minimum 
$h$ allowed for p-model \cite{MS87} She-Leveque \cite{SL94}, 
respectively, which preclude
saturation; whereas $D(h=0) \approx 3-2.1 = 0.9$ marks saturation 
for transverse exponents at $\zeta_\infty^{tr} \approx 2.1$.
}
\label{fig:dh}
\end{figure}

The Lagrangian exponents can be computed from Eq.~\eqref{eq:zlag} by using
Eulerian multifractal spectrum $D(h)$ from Eq.~\eqref{eq:zeul}. 
The $D(h)$ corresponding to the Eulerian multifractal models shown 
in Fig.~\ref{fig:eul} are plotted in Fig.~\ref{fig:dh}.
They are obtained from $\zeta_p$
by taking a Legendre transform to invert the
relations \cite{Frisch95}, giving
\begin{align}
    D(h) = \inf_p \left[ ph + 3 - \zeta_p \right].
\label{eq:dh}
\end{align}
For reference, the 
$D(h)$ for She-Leveque model is \cite{SL94}
\begin{align}
D(h) = 1 + c_1 (h - h^*) - c_2 (h - h^*) \log (h-h^*)
\end{align}
where $h^*=1/9$, 
$c_1 = c_2 (1 + \log \log \gamma - \log \gamma)$
and $c_2 = 3/\log\gamma$, with $\gamma=3/2$. That for the Sreenivasan-Yakhot result 
of $\zeta_p = \zeta_\infty p/(p+ \beta)$ \cite{SY2021} is 
\begin{align}
D(h) = 3 - \zeta_\infty -  \beta h + 2 \sqrt{ \zeta_\infty \beta h}
\end{align}
where $\zeta_\infty \approx 7.3$ and $\beta=3\zeta_\infty-3$.
The result for p-model can be found in \cite{MS87}.

In Fig.~\ref{fig:dh}, 
in addition to the $D(h)$ from 
these known Eulerian cases, we also 
utilize Eq.~\eqref{eq:dh} 
to numerically obtain the $D(h)$ for transverse exponents
(with $\zeta_p^{tr} \approx 2.1$ for $p \ge 10$, as shown
in Fig.~\ref{fig:eul}). 
Note, since the $D(h)$ for $\zeta_p^{tr}$
is obtained numerically, the inversion formula in Eq.~\ref{eq:dh}
can only provide the concave hull \cite{Frisch95}---which 
is what we plot in Fig.~\ref{fig:dh}.
The saturation value of exponents is reflected 
in the corresponding
$D(h)$ curve for $h=0$, 
as $D(0)  = 3 - \zeta_\infty$
($\approx 0.9$ for $\zeta_\infty^{tr} \approx 2.1$).
Note, $h<0$ is not allowed in the multifractal framework
\cite{Frisch95}; the p-model and She-Leveque results 
respectively correspond to 
$h_{\rm min} = \frac{1}{3}\log_2 (0.7) \approx 0.172$ \cite{MS87}
and $h_{\rm min} = h^* = \frac{1}{9}$ \cite{SL94}, which preclude saturation. 
The Sreenivasan-Yakhot result \cite{SY2021}
predicts saturation for longitudinal 
exponents (at $\zeta_\infty \approx 7.3$,
giving $D(0) = 3 - 7.3 = -4.3$ (not shown in 
in Fig.~\ref{fig:dh}).

\paragraph*{Lagrangian exponents from the transverse multifractal spectrum:}
As we saw, none of the multifractal predictions for Lagrangian exponents 
using Eulerian longitudinal exponents agree with the data.
In contrast, the prediction 
corresponding to transverse Eulerian exponent
(green dot-dashed line in Fig.~\ref{fig:lag}) 
closely follows the measured results, 
particularly capturing the saturation at high orders. 
Note, the predicted saturation value $\zeta_\infty^L \approx 2.1$, 
is the same for both transverse Eulerian and Lagrangian exponents,
The actual Lagrangian data
saturate at a very slightly smaller value. 
We believe this minor difference
(of only 5\%)
stems from the fact that 
even at $\re=1300$, 
the temporal inertial-range
is underdeveloped, and
the intermittency-free result
of $\zeta_2^L=1$ is not unambiguously realized. 
Since 
Lagrangian exponents shown in Fig.~\ref{fig:lag} 
are extracted as ratios $\zeta_p^L/\zeta_2^L$,
this minor discrepancy in the saturation values
could be explained by 
small departures from the expectation of $\zeta_2^L=1$.
Given this and also possible statistical uncertainties
(at highest orders), the close correspondence
between the transverse Eulerian exponents
and Lagrangian exponents is quite remarkable.

It is worth noting that 
Lagrangian exponents saturate 
for slightly smaller $p$ than for transverse Eulerian exponents.
This readily follows from Eqs.~\eqref{eq:zeul}-\eqref{eq:zlag} 
as a kinematic effect.
For Eulerian exponents,
$\zeta_3 = 1$ is exact, 
corresponding to $h\approx \frac{1}{3}$, $D(h) \approx 3$, which conforms
to the intermittency-free K41 
result \cite{Frisch95}. 
This gives $\zeta_2^L = 1$
as the corresponding Lagrangian result
for $h\approx \frac{1}{3}$, $D(h) \approx 3$.
This argument can be extended to higher orders
to show that
Lagrangian exponents at order $p$ correspond to
transverse exponents at order $3p/2$.
Thus, it follows that 
Lagrangian exponents saturate 
at smaller $p$.
A similar  correspondence
can also be provided for other Lagrangian statistics,
for instance, the second-moment of 
acceleration (the temporal velocity-gradient) 
corresponds to the third-moment of spatial velocity-gradients
\cite{MY.II, BS_PRL_2022}.

\paragraph{Discussion:} 
Two significant results emerge from our work:
(a) scaling exponents saturate for both transverse Eulerian and
Lagrangian structure functions; and 
(b) the saturation of Lagrangian exponents
is characterized solely by the transverse Eulerian
exponents (and not the longitudinal, as previously believed).
Given that the transverse exponents are smaller for large $p$, this
seems reasonable from the steepest-descent argument \cite{Frisch95}.

The saturation of scaling exponents is
extreme form of anomalous behavior, but is
not uncommon; it holds for 
Burgers equation \cite{bec07},
passive scalar turbulence 
\cite{RK94,iyer18, BCSY2021b}. 
However, its prevalence in velocity field 
has become apparent only recently
\cite{iyer20, SY2021}. The theory of \cite{SY2021} 
predicts that Eulerian longitudinal exponents saturate as well, 
although at very high moment orders that cannot be yet validated. 
In contrast, both transverse Eulerian exponents 
and Lagrangian exponents saturate and
at the same value of $\approx 2$. 
Further, using a simple physical correspondence based on 
frozen flow hypothesis, they are
related through the same multifractal spectrum
(which differs from known spectrum for longitudinal
Eulerian exponents).
Interestingly, the saturation exponent of 2 implies a
fractal co-dimension of 1 \cite{Frisch95, Sreeni97}, suggesting 
that the saturation likely comes from localized (very) thin vortex
filaments, which are known to be
prevalent at the smallest scales \cite{Jimenez93, Ishihara09, BPBY2019}.



Our results also bring forth some important questions. 
First is the extension of the multifractals
from inertial- to dissipative-range, i.e.,
describing the scaling of velocity-gradients.
Such an extension relies on the
phenomenological criterion 
that the local Reynolds number,
describing the dissipative cutoff,
is unity, i.e., 
$\delta u_r r/\nu = 1$ \cite{Paladin87, Frisch95, SY2021}. 
As highlighted in recent works \cite{BPBY2019, BP2022},
this is valid for longitudinal
increments, but not
for transverse increments, essentially because
of how vorticity and strain-rate interact in turbulence.
It can thus be expected that the extension 
of multifractals to dissipation-range works
for longitudinal velocity-gradients,
but not for transverse velocity-gradients.
Since the current results suggest that
Lagrangian intermittency 
is linked to transverse Eulerian intermittency,
it follows that the extension to acceleration statistics
would be an issue, as confirmed by our recent studies \cite{BS_PRL_2022, BS2023}.
In addition, acceleration components are
strongly correlated in turbulence \cite{tsinober01, BS2023}, which is a feature of  
Navier-Stokes dynamics that is not accounted for by multifractals.

A second question concerns the meaning of universality given the longitudinal 
and transverse exponents behave differently. One strategy could be to consider 
a joint multifractal spectrum for longitudinal and
transverse increments.  
It might be possible to set appropriate
conditions on both to enable the inertial-range universality 
and the transition from the inertial- to dissipation-range.
Essentially, addressing the discrepancy between longitudinal 
and transverse intermittency presents a critical and pressing problem 
in turbulence theory.

\begin{acknowledgements}
\paragraph*{Acknowledgments:}
We gratefully acknowledge discussions with Victor Yakhot and
sustained collaboration with P.K. Yeung. 
We also gratefully acknowledge
the Gauss Centre for Supercomputing 
e.V. (www.gauss-centre.eu) for providing computing time 
on the supercomputers 
JUQUEEN and JUWELS at J\"ulich Supercomputing Centre (JSC),
where the simulations reported in this
paper were primarily performed.
Computations were also supported partially by the supercomputing
resources under the Blue Water project 
at the National Center for Supercomputing Applications 
at the University of Illinois (Urbana-Champaign).
\end{acknowledgements}


%

\end{document}